\documentclass[proceedings]{JHEP} 

\usepackage[latin1]{inputenc}

\usepackage{rotating}

\usepackage{exscale}

\usepackage{epsfig}

\newcommand{\bea}{\begin{eqnarray}}
\newcommand{\eea}{\end{eqnarray}}

\newcommand{\be}{\begin{equation}}
\newcommand{\ee}{\end  {equation}}
\newcommand{\beq}{\begin{eqnarray}}
\newcommand{\eeq}{\end{eqnarray}}
\newcommand{\pa}{\partial}
\newcommand{\half}{{1\over 2}}

\newtheorem{theorem}{Theorem}

\newcommand{\pat}{\partial_t}
\newcommand{\px}{\partial_x}
\newcommand{\crr}{\nonumber\\}
\newcommand{\eel}[1]{\label{#1}\end{equation}}
\newcommand{\eeal}[1]{\label{#1}\end{eqnarray}}
\newcommand{\baq}{\begin{equation}\begin{array}{rcl}}
\newcommand{\eaq}{\end{array}\end{equation}}
\newcommand{\eaql}[1]{\end{array}\label{#1}\end{equation}}
\newcommand{\beac}{\begin{equation}\begin{array}{rcl}}
\newcommand{\eeacn}[1]{\end{array}\label{#1}\end{equation}}
\newcommand{\ba}{\begin{array}}
\newcommand{\ea}{\end{array}}
\newcommand{\non}{\nonumber \\}

\renewcommand{\a}{\alpha}

\newcommand{\adss}{$AdS_5\times S^5$\ }

\conference{Non-perturbative Quantum Effects  2000}

\title{ Stringy Confining Wilson Loops}

\author{Jacob Sonnenschein \thanks{
This talk is based on works done in collaboration
with A. Armoni, A. Brandhuber, E. Fuchs, N. Itzhaki, Y. Kinar, A. Loevy, E. Schreiber,
N. Weiss and S. Yankielowicz .}\\
        School of Physics and Astronomy\\
Beverly and Raymond Sacler Faculty of exact Sciences\\
Tel Aviv  University, Ramat Aviv, Tel Aviv 69978, Israel 
        E-mail: \email{cobi@ccsg.tau.ac.il}}

\abstract{
The extraction of Wilson loops of confining gauge systems from
their SUGRA (string) duals is reviewed. I start with describing the basic
classical setup. A theorem that determines the classical values of the
loops associated with a generalized background is derived. In particular
sufficient conditions for confining behavior are stated .
 I  then introduce quadratic
quantum fluctuations around the classical configurations.
I  discuss in details the following models of confining behavior: (i)
Strings in flat space-time, (ii) $AdS_5$ black hole and its correspondence
with pure YM theory in three dimensions. 
In particular an attractive Luscher term
is shown to be the outcome of the quantum fluctuations.
(iii) Type 0 string model (iv) The   Polchinski Strassler $N=1*$ model. In the latter case we show that SUGRA
alone is not enough to get the correct nature of the loops, and only
by incorporating the worldvolume phenomena of the five branes a coherent
qualitative picture can be derived.   }


\begin{document}

\newpage
\begin{center}
\fbox{
\bf A brief reminder on Wilson loops}
\end{center}
\begin{itemize}
\item
In $SU(N)$ gauge theories one defines the following set of gauge invarinat
operators
\begin{center}
\fbox{
$ W(C) = {1\over N} Tr P e^{\oint_C A_\mu \dot x^\mu (\tau) d\tau} $}
\end{center}
where $C$ is some contour.


\item 
In this talk I restrict myself to $C$ which is an infinite strip as is shown
in figure 1. 


\item The quark anti-quark potential $E(L)$  can be extracted from the 
infinite strip Wilson loop  as follows

\begin{center}
\fbox{
$ \langle W(C)\rangle  = A(L)   e^{ -TE(L)}. $}
\end{center}
\item The natural  (bosonic) stringy  candidate for the Wilson loop 
(which obeys the loop equation) is

\begin{center}
\fbox{
$ \langle W(C)\rangle \sim    e^{ -S^{ren}_{NG}} $}
\end{center}
 where $S^{ren}_{NG}$ is the renormalized NG action\cite{Mal2,Rey,DGO},  
which is the world sheet area of the string. 
The renormalization has a simple physical intepretation of 
subtracting the quark masses.
\begin{figure}[h!]
\begin{center}
\resizebox{0.4\textwidth}{!}{\includegraphics{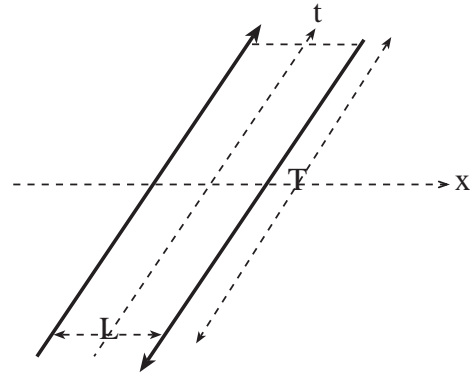}}
\end{center}
\caption{The basic setup of the Wilson loop}
\end{figure}

\end{itemize} 


\begin{center}
\fbox{
\bf Stringy Wilson loop- general setup }
\end{center}
We now introduce the basic setup  which will serve us in analysing 
Wilson loops of   various string backgrounds\cite{KSS2}.
Consider a 10d space-time metric 
\bea
 ds^2 &=& -G_{00}(s) dt^2 + G_{x_{||}x_{||}}(s) dx^2_{||} \non
&+& G_{ss}(s) ds^2
 + G_{x_{T}x_{T}}(s) dx^2_{T} \non
\eea
where  
$x_{||}$- $p$ space coordinates on a $D_p$ brane 

$s$ and  $x_T$ are the transverse coordinates 

The corresponding Nambu-Goto action is 
$$
S_{NG} =\int d\sigma d\tau \sqrt{det[\pa_\alpha  x^\mu \pa_\beta x^\nu
G_{\mu\nu}]}. \crr
$$
Upon using $\tau=t$ and  $\sigma=x$, where  $x$ is  
the space coordinate of the loop which is 
one of the $x_{||}$,
    the  action 
for a static configuration  reduces  to  
$$
S_{NG}= T \cdot \int dx\sqrt{ f^2(s(x)) + g^2(s(x)) (\pa_x s)^2} \crr
$$ 
where
\begin{eqnarray*}
f^2(s(x)) & \equiv & G_{00}(s(x))G_{xx}(s(x)) \crr
g^2(s(x)) & \equiv & G_{00}(s(x))G_{ss}(s(x)) \crr
\end{eqnarray*}
and $T$ is the time interval.

 The equation of motion  (geodesic line) 

$$
\label{sofx}
\frac{d s}{d x} = \pm \frac{f(s)}{g(s)} \cdot
\frac{\sqrt{f^2(s)-f^2(s_0)}}{f(s_0)}
$$


 A static string configuration connecting quarks which are  separated  
by a distance
$$
\label{lgeneral}
L = \int dx = 
2 \int_{s_0}^\infty \frac{g(s)}{f(s)} \frac{f(s_0)}{\sqrt{f^2(s)-f^2(s_0)}} ds 
$$
To have a finite separation distance the slope $\frac{d s}{d x}$
has to diverge on the boundary. 

The NG action   
and  corresponding  energy  $E={S_{NG}\over T}$ are 
divergent.  
The action  is renormalized  by\cite{Mal2} 

(i) regularizing the integral $\int^\infty\rightarrow \int^{s_{max}}$ 

(ii)
subtracting  the quark masses
$$
m_q = \int_0^{s_{max}} g(s) ds
$$

So that the renormalized quark anti-quark potential is

\begin{center}
\fbox{
$E  =  f(s_0) \cdot L  $} 
\begin{eqnarray*}
 &+& 2 \int_{s_0}^{\infty} \frac{g(s)}{f(s)}  
( \sqrt{f^2(s)-f^2(s_0)} - f(s)  ) ds \non
&-& 2 \int_0^{s_0} g(s) ds \non
\end{eqnarray*}
\end{center}

The behavior of the potential is determined by  the following theorem
\cite{KSS2}. 
\begin{theorem}
\label{main}
Let $S_{NG}$ be  the NG action defined above, with functions $f(s),g(s)$ such
that:
\begin{enumerate}
\item \label{smoothf} $f(s)$ is analytic   for $0 < s < \infty$.
At $s = 0$, ( we take here that the minimum of $f$ is at $s=0$ ) its expansion is:
$$
\label{expansion}
f(s) = f(0) + a_k s^k + O(s^{k+1})
$$ 
with $k > 0 \;,\; a_k > 0$.
\item \label{smoothg} $g(s)$ is smooth for $0 < s < \infty$. At $s = 0$,
its expansion is:
$$
g(s) = b_j s^j + O(s^{j+1})
$$
with $j > -1 \;,\; b_j > 0$.
\item \label{positive} $f(s),g(s) \ge 0$ for $0 \le s < \infty$.
\item \label{increasing} $f'(s) > 0$ for $0 < s < \infty$.
\item \label{inftyint} $\int^\infty g(s)/f^2(s) ds < \infty$.

\end{enumerate}
Then for (large enough) $L$ there will be an even geodesic line asymptoting
from both sides to $s = \infty$, and $x = \pm L/2$.
The associated potential  is 
\begin{enumerate}
\item \label{conf}   if $f(0) > 0$, then
\begin{enumerate}
\item if $k = 2(j+1)$, 
\vskip 1cm
\begin{center}
\fbox{
$E = f(0) \cdot L -2 \kappa + 
                                    O((\log L)^\beta  e^{-\alpha L})$}
\end{center}
\item if $k > 2(j+1)$, 
\begin{center}
\fbox{
$E = f(0) \cdot L -2 \kappa - d \cdot L^{-\frac{k+2(j+1)}{k-2(j+1)}} 
+ O(L^\gamma ) $.}
\end{center}
\end{enumerate} where $\gamma={-\frac{k+2(j+1)}{k-2(j+1)} - \frac{1}{k/2-j}}$
 and $\beta$ and $\kappa$, $\alpha$  $d$ and   $C_{n,m}$ are
positive constants determined by the string configuration.

In particular, there is 
\begin{center}
\fbox{

linear confinement}

\end{center}

\item \label{noconf} if $f(0) = 0$, then if $k > j+1$,
\begin{center}
\fbox{
$E = -d' \cdot L^{-\frac{j+1}{k-j-1}}
\ \ \  + O(L^{{\gamma}'})$} 
\end{center}
where $\gamma'={-\frac{j+1}{k-j-1} 
- \frac{2k-j-1}{(2k-j)(k-j-1)}}$ and 
  $d'$ is a  coefficient determined by the classical configuration.

In particular, 
\begin{center}
\fbox{

there is no  confinement}

\end{center}




As a corollary of this theorem\cite{KSS2} we  find that 
if one of the following two conditions is obeyed :

(i) $f$ has a minimum at $s_{min}$
and $f(s_{min})> 0$,

(ii)  $g$ diverges at $s_{div}$ and
 $f(s_{div})> 0$ 
then  the corresponding Wilson loop  {\bf confines}.

 \begin{figure}[h!]
\begin{center}
 \resizebox{8cm}{!}{
   \includegraphics{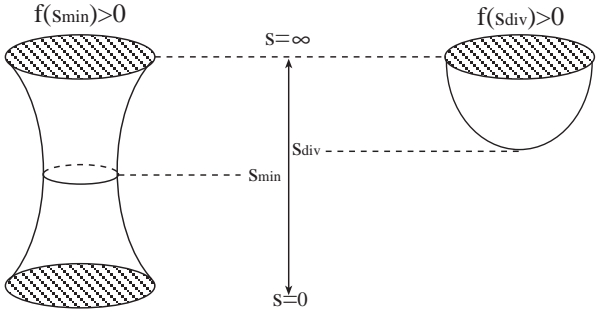}
   }
\end{center}
\caption{Sufficient conditions for confinement.  
 }
\end{figure}

\end{enumerate}
\end{theorem}

%
\begin{center}
\fbox{
\bf Quantum fluctuations
}
\end{center}

Introduce  quantum fluctuations around the classical configuration

$$
x^\mu(\sigma,\tau) = x^\mu_{cl}(\sigma,\tau) + \xi^\mu(\sigma,\tau)
$$

The quantum corrections  to the Wilson line  to quadratic order is 
\cite{KSSW}
$$
        \langle W\rangle = {\rm e}^{-TE_{cl}(L)}~
        \int \prod_a d\xi_a \exp\left(-\int d^2 \sigma \sum_a \xi^a {\cal O}^a\xi^a \right)
$$
where $\xi^a$ are the fluctuations left after  gauge fixing. 
The corresponding correction to the free energy is 
$$
F_B = -\log {\cal{Z}}_{(2)}= -\sum_a\half \log \det {\cal O}_a\non
$$

\begin{center}
{
\bf general form of the bosonic determinant
}
\end{center}

In the $\sigma=u$ gauge ( after a change of variables) 
the free energy is given by 
\bea 
\label{FrEn}
       F_B &=& - \half \log {\det {\cal O}_{x}}
 - {(p-1)\over 2} \log {\det {\cal O}_{x_{II}}} \non
&-& {(8-p)\over 2} \log {\det {\cal O}_{x_T}}\non
\eea
where 
\beq
\label {QuadOp} 
\hat{\cal O}_x &=& \left[\px\left((1-{f^2(u_0) \over f^2(u_{cl})})
\px\right) + {G_{xx}(u_{cl}) \over G_{tt}(u_{cl})} ({f^2(u_{cl}) \over f^2(u_0)}-1)
\pat^2 \right] \nonumber \\
\hat{\cal O}_{x_{II}} &=&  \left[\px\left({G_{y_i
y_i}(u_{cl}) \over G_{xx}(u_{cl})} \px\right) + {G_{y_i y_i}(u_{cl}) \over G_{tt}(u_{cl})} 
{f^2(u_{cl}) \over f^2(u_0)} \pat^2 \right]   \nonumber \\
\hat{\cal O}_{x_T} &=& ... \nonumber \\
\eeq
where $\hat{\cal O}= {2\over f(u_0)}{\cal O}$
and the boundary conditions are $\hat\xi(-L/2,t) = \hat\xi(L/2,t)=0$.
The fermionic fluctuations will be discussed for each model 
separately.

{\it model 1}

\begin{center}
\fbox{
\bf Wilson loop from string in flat space-time 
}
\end{center}

\begin{itemize}
\item Consider the bosonic string in flat space-time with the boundary conditions
$$x(\sigma=0)=0 \qquad  x(\sigma=\pi)=L $$

The static NG action  takes the form
$$ S_{NG}= T_{st} \int dx  \sqrt{1+ (\pa_x u)^2} $$
where $T_{st}={1\over 2\pi \alpha'}$.

The classical static  configurations  are flat, 
$u'=0$


The quark anti-quark potential  that follows from  the NG action is
$$ V(L) =  T_{st} L $$
Thus,  the  classical stringy ``Wilson loop''   implies  
  a linear confining potential.


{\bf Bosonic fluctuations in flat space-time }

Let us turn now on  quantum fluctuations\cite{KSSW}. 

The  action in this case takes this simple  form 
$$
        S_{(2)}=\half\int d\sigma d\tau
\sum_{i=1}^{D-2}\left[ 
        \left(\partial_\sigma\xi_i\right)^2 +
        \left(\partial_\tau\xi_i\right)^2
        \right]
$$
The  corresponding eigenvalues are 
$$
       \lambda_{n,m} = ({{n \pi \over L}})^2 + ({{m \pi} \over T})^2
$$
and the free energy is given by
$$
       -\frac{2}{D-2} F_B =   \log \prod_{n m} \lambda _{n,m} = 
 T {\pi \over 2L} \sum_n n + O(L)
$$

Regulating this result  using Riemann $\zeta$ 
function we find
that the  quantum correction to the linear quark anti-quark potential 
is

\begin{center}
\fbox{
$
      \Delta V(L) = - {1 \over T}  F_B 
= -(D-2) {\pi \over 24} \cdot {1 \over L}$
}
\end{center}

which is the so-called {\bf  L\"{u}scher term}\cite{Luscher}.   

\begin{center}
{\bf The fermionic fluctuations in flat space-time}
\end{center}
\item 
We use here the Green Schwarz  action since in  cases 
which will discuss later  the NSR action is not known.


  .

\item The fermioic part of the $\kappa$ gauged fixed  GS-action is
$$
    S_F^{flat} = 2 i  \int d\sigma d \tau  \bar \psi  \Gamma^i \pa_i \psi 
$$
where $\psi$ is a Weyl-Majorana spinor, $\Gamma^i$ are 
the SO(1,9) gamma matrices, $i,j = 1,2$ and we explicitly considered a flat classical string.

Thus the fermionic operator is
$$ 
    \hat{\cal O}_F=D_F = \Gamma^i \pa_i 
$$
and squaring it we get 
$$(\hat{\cal O}_F)^2=\Delta = \px^2 - \pat^2$$  
The total free energy  is 
$$
    F = 8 \times \left(- \half \log \det \Delta + \log \det D_F \right) = 0 
$$
since for D=10, we have 8 transverse coordinates 
and 8 components of the unfixed  Weyl-Majorana spinor.

\item Thus, 
in flat space-time  the classical 
stringy Wilosn loop 
is not corrected by   quadratic quantum fluctuations.


\begin{center}
{\bf Can the Wilson line be evaluated exactly?}\end{center}

Let us consider the bosonic string with the boundary
conditions given above\cite{Arvis}.
\item The energy 
of any string state is given by 
$$
E^2=P^2 +4(L_0-a) =(LT_{st})^2 +4(L_0-a) \nonumber\\
$$
Thus for the 
 lowest {\bf tachyonic} state $(L_0=0)$ it  is given by
$$
E^2=P^2 + m_{tach}{^2}=(L T_{st})^2 -T_{st}{\pi(D-2)\over 12} \nonumber\\
$$
If we assign the potential with this energy we have  
\begin{center}
\fbox
{$
V(L) =T_{st} L\sqrt{1- {\pi (D-2)\over 12}{1\over T_{st}L^2}} 
 $}
\end{center}

which can be expanded 
 $$\sim T_{st}L -\pi {(D-2)\over 24}{1\over L}+ ...
 $$
Thus this expansion yields the  leading confining behavior as well as the 
Luscher  quadratic fluctuation term. 
\item Moreover 
for a bosonic string in Flat space-time O.Alvarez\cite{Alvarez}
 showed that in the large
$D$ limit
$$ D\rightarrow \infty \qquad {\pi \over 24 T_{st}L^2}\rightarrow 0\qquad
{D\pi \over 24 T_{st}L^2}\rightarrow finite$$
using  the variables $\sigma_{\alpha\beta}=\pa_\alpha x^\mu
\pa_\beta x_\mu$
the exact effective action is  
$$S^{exact}_{NG}=T_{st} L\sqrt{1- {\pi (D-2)\over 12}{1\over T_{st} L^2}} $$
\item for $L<\sqrt{{ \pi(D-2)\over 12} {1\over T_{st}}}$ this approach fails.


\end{itemize}

\begin{center}
\fbox{
\bf The Ads balck hole and pure YM theory in 3d }
\end{center}

\begin{itemize}
\item Can we detect the confining nature of pure YM theory

\item In   Field theory $YM_3$ can be reached from ${\cal N}=4$ SYM by:

(i) Compactifying the Euclidean time direction ( introducing temperature)

(ii) Imposing anitperiodic boundary conditions.

\item Recall that due to such boundary conditions

{\bf supersymmetry is broken} , 

and in the case of the ${\cal N}=4$ SYM the fermions and scalars 
become massive\cite{Witads2}
 $$m_{fermions}\sim T \qquad m_{scalars}\sim g_{YM4}^2 T$$
\item  In the limit of $T\rightarrow \infty$  the 
Euclidean  4d theory turns into 
a 3d theory, and since the fermions and scalars decouple it is  
pure $YM_3$  with the coupling 

$$
g_{YM_3}^2 = g_{YM_4}^2 T ~.
$$

\item In the SUGRA picture the introduction of temperature translates
into the use of near extremal \adss solution

 \item The metric of near extremal D3 branes in the large N limit is 
\bea\label{metric}
&& {ds^2\over \a'}  =  \frac{U^2}{R^2} 
[ - f(U) dt^2 + dx_i^2]  \non
&&+  R^2 f(U)^{-1} 
    \frac{dU^2}{U^2}
+ R^2 d\Omega_5^2  \non
&& f(U)  =  1 - U^4_T/U^4 \non
&& R^2 = \sqrt{4 \pi g N} ,~~~~~~~~~~~ U_T^4 = 
\frac{2^7}{3} \pi^4 g^2 \mu ~,\nonumber
\eea
where $\mu$ is the energy density.


\item The idea is thus to 
consider the Wilson loop 
along two space directions for  the case of the 
near extremal D3 brane solution\cite{BISY2}.


\item  We take   $Y\rightarrow\infty$  
which will be the 3d Euclidean time direction and the other
direction, $L$, to be finite.
 

\item For such a setup we have
$$ f^2= ({U\over R})^4 \qquad  g^2= {1\over (1-({U_T\over U})^4)} $$
since $g^2$ diverges at $U=U_T$ we must have confinement.

\item Indeed, let us insert the metric of above to the NG action 
\be
S = \frac{Y}{2\pi}\int dx \sqrt{\frac{U^4}{R^4}+\frac{(\partial_x
    U)^2}{1-U_T^4/U^4}}\non
\ee
\item The distance between the quark and the anti-quark is
\be\label{jk}
L = 2 \frac{R^2}{U_0} \int_{1}^{\infty}
\frac{dy}{\sqrt{(y^4-1)(y^4-\lambda)}},\non 
\ee
where $\lambda =U_T^4/U_0^4 <1$ and
 $U_0$ is the minimal value of $U$.

\item The energy is 
\beq\label{kj}
E&=&\frac{U_0}{\pi}\int_{1}^{\infty}dy\left( \frac{
    y^4}{\sqrt{(y^4-1)(y^4-\lambda)}}-1\right)\\
 &+&\frac{U_T-U_0}{\pi},\non
\eeq
\item Notice that in the limit $U_0\rightarrow  
U_T$ ($\lambda\rightarrow 1$) we get $L\rightarrow\infty$. 
In this limit the main contribution to the integrals of $L$ 
and  $E$  comes from the region near $y=1$.

\item Therefore, we get for large $L$
\begin{center}
\fbox{
$ E=T_{QCD} L  $
}
\end{center}
\begin{center}
\fbox{
$ T_{QCD} = \frac{\pi}{2} R^2 T^2, $
}
\end{center}

\item Notice that the string tension diverges in the SUGRA limit
since 
$$ R=\sqrt{g_{YM}^2N}\rightarrow \infty  
\qquad T\rightarrow \infty$$ 

\item If there are no phase transitions in going from the SUGRA limit
$ R=\rightarrow \infty $ to the  full stringy description 
of $YM_3$ then indeed the latter predicts {\bf confinement}







\begin{center}
\fbox{
{\bf The  determinant   for ``confining scenarios"}}
\end{center}

\item
Let us consider now the quantum fluctuations in 
this SUGRA  setup which is dual to the pure YM theory
in 3d ( AdS black hole in the $T\rightarrow\infty$ limit)\cite{KSSW}.  
Now we have 
\beq
f(u) & = & u^2/R^2 \non
g(u) & = & (1 - (\frac{u_T}{u})^4)^{-1/2}\non
\eeq
\item In the large $L$ limit the classical string is very flat
with $u\sim u_0 $. In fact as $L$ grows $u_0\rightarrow u_T$.
In this limit  
\beq
\hat{\cal O}_y & \longrightarrow & \frac{u_T^2}{2} \left[ \partial_x^2 +
  \partial_t^2 \right] \nonumber\\
\hat{\cal O}_\theta & \longrightarrow & \frac{R^2}{2} \left[ \partial_x^2 +
  \partial_t^2 \right] \nonumber\\
\hat{\cal O}_z & \longrightarrow & 2 u_T^2 e^{-2 u_T L} \left[ \partial_x^2 +
  \partial_t^2 \right] \nonumber\\
\hat{\cal O}_n & \longrightarrow & \left[ \frac{4 u_T^2}{2R^4} +
  \frac{1}{2} \partial_x^2 + \frac{1}{2} \partial_t^2 \right]\nonumber\\
\eeq
\item We see that the operators for transverse fluctuations, 
$\hat{\cal O}_y$, $\hat{\cal O}_z$, 
turn out to be simply the Laplacian in flat spacetime, multiplied 
by overall
factors, which are irrelevant.  Therefore, the
transverse fluctuations yield the
standard L\"uscher term  proportional to $1/L$. 
\item The longitudinal normal fluctuations give rise to an operator 
$\hat{\cal O}_n$ corresponding to a scalar field with mass $2 u_T/R^2 = \alpha$.
Such a field contributes a Yukawa like 
 term $$\approx -\frac{\sqrt{\alpha}e^{-\alpha L}}{\sqrt{L}}$$
to the potential. 
\item Thus, altogether there are 7 Luscher type modes
 and one massive mode.  
\item 
Now we have to turn on the fermionic fluctuations. 
Had the fermionic modes been those of flat space-time 
then the total coefficient
in front of the Luscher term would have been
$+8-7=+1$, namely, 
a repulsive Culomb like potential\cite{GriOle,DorPer}. 
This contradicts gauge dynamics\cite{Bachas}. 

\item  There is   a GS formulation for the \adss background\cite{MetTse,Pesando
KalTse},
but the analog for the Ads black hole has not been written down.   
Nevertheless, we argue that the coupling of the fermion to the RR
field is the same as for the extremal \adss background.
(Since the dilaton, the RR field and $det(G_{\mu\nu}$ are unchanged)  
\item  For that case we found that in the large $L$ limit 
the square of the fermionic operator is 
$$ \hat{\cal O}^2_\psi =\frac{u_T^2}{2} \left[ \partial_x^2 +
  \partial_t^2 +({U_T\over R^2})^2\right ] $$

\item If the assumption about the coupling to the RR is correct,
the quark anti-quark potential is corrected
by an  attractive Luscher term 

$$ -7{\pi\over 24}{1\over L} $$

\end{itemize}


{\it model 3.} 

\begin{center}
\fbox{
\bf Wilson loops in type 0 string  theory
}
\end{center}
\begin{itemize}
\item
{\bf What is type 0 string}

Type 0 string is supersymmetric on the world sheet but not in space-time
due to a non-chiral GSO projection. The type $0_A$ and type $0_B$ differ from 
the  type $II_A$ and type $II_B$
(i) No space-time fermions,
(ii) Doubling of the RR fields,
(iii) Tachyons.  
\item
A type 0 model can be made consistent only provided
(i) The Tachyon $m_{tach}^2$ can be shifted to $m_{tach}^2>{c\over R^2}$
(ii) No dilaton ( and possible other massless fields) tadpoles
(iii) The low energy effective theory is reliable if 
$g_{st}<<1 \qquad {\cal R}<<1$ where ${\cal R}$ is the scalar curvature 
in the string frame.

\item 
The Wilson loops were discussed both in the critical string and 
in Polyakov's   non-critical string  model.
\item 
The equations of motion of the  low energy effective theory guarantee that 
\cite{AFS}
$$ \pa^2_s f(s) \geq 0$$

\item 
The interpretation of an IR and UV domains may be in terms of the  structure
of the Wilson line is  as is shown in figure 4. 

\begin{figure}
\begin{center}
 \resizebox{8cm}{!}{
   \includegraphics{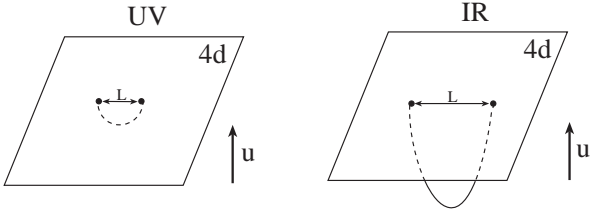}
   }
\end{center}
\caption{The IR and UV regimes}

\end{figure}
so that the large $u$ regime corresponds to the gauge theory UV regime and 
the small  $u$ regime to the IR. 
\item In the  IR the generic solution has 
$$ \pa_s f(s) = 0 \qquad with\ \qquad f(s_{min})\neq 0$$
  So that generically the solution in the IR admits a linear confinement 
behavior. 
\item 
This can also be verified from  arguments based on the  5d bulk theory 
and in particular also from the  screening nature of the 
't Hooft loop\cite{gppz}. 

\item  In the UV a fixed point in  the form of $AdS_5\times S^5$ was 
observed.  
Moreover around the fixed point $f\sim log L$ so that
it was argued  that\cite{Minahan} 
$$\Delta V_1\sim {1\over log{L_0\over L}}\ {1\over L}$$
It was further found that the  higher order correction 
produces a WIlson line\cite{KleTse2} 
$$\Delta V_2\sim {1\over  (log{L_0\over L}-c log\ log{L_0\over L})}\ {1\over L}$$
which resembles the 2 loop correction in  the gauge  theory picture.  
Note however that in the UV generically the curvature in the string frame is
not negligible and thus the assertions have to be made with a grain of
salt.  

\end{itemize}

{\it model 4.} 

\begin{center}
\fbox{
\bf Loops in the Polchinski Strassler $N={1}^*$ theory
}
\end{center}

We start with  a brief review of the theory\cite{PolStr} 

{\bf The field theory picture}
\begin{itemize}
\item Perturb the ${\cal N}=4$  $SU(N)$ 
SYM theory by adding a mass term to the 
superpotential.
\begin{center}
\fbox{
$ W+\Delta W = {2\sqrt{2}\over g^2_{YM}}tr\left( [\phi_1,
\phi_2]\phi_3\right )  + {m\over g^2_{YM}}\sum_{i=1}^3\phi_i^2$
      }
\end{center}

where $\phi_i$ are the 3 complex scalars of the ${\cal N}=4$ .

\item The {\it classical vacua} are given by $N$ dimensional  reducible reps. of 
$SU(2)$ since the equation of motion is
$$ [\phi_i,\phi_j]={-m\over \sqrt{2}}\epsilon_{ijk} \phi_k$$

\item The {\it quantum vacua} correspond to  order $N$ subgroups 
of 
$$Z_N\times Z_N$$
\item The perturbation also turns on a mass term to 3 out of the 4 
Weyl fermions
$$m^{\alpha\beta}\lambda_\alpha\lambda_\beta +h.c $$

 
{\bf The SUGRA picture}

\item The perturbing fermionic   mass term,
$\bar 10$ of $SU(4)$ corresponds to 
turning on a {\it magnetic 3 form}  obeying 
$$ *_6T= iT $$
\item Explicitly, denoting the 6 transverse coordinates by  3 complex
coordinates $z_i$ $i=1,2,3$,  the 3 form  T is 
$$ T_3= m[dz_1\wedge d\bar z_2\wedge d\bar z_3 + 
d\bar z_1\wedge d  z_2\wedge d\bar dz_3  + d\bar z_1\wedge d\bar z_2\wedge d
dz_3 ] $$
\item Further breaking to $N=0^*$ can be achieved by adding a term
$ m'dz_1\wedge d z_2\wedge  dz_3$ to $T$.

\item The coupling of the $D_3$ branes  to the magnetic 3 form 
produces via the {\it Myers Polarization } mechanism\cite{Myers}  {\it five branes}
that wrap $S^2$.

\item The metric of the $N=1^*$ models takes the form
\bea
 ds^2 &=& Z_x^{-1/2} \eta_{\mu\nu} dx^\mu dx^\nu \non
&+&  Z_y^{1/2} (dy^2 + y^2
d\Omega_y^2 + dw^2) + Z_\Omega^{1/2} w^2 
d\Omega_w^2\non 
\eea
denoting by 
$Z_0= {R^4\over \rho_+^2\rho_-^2}\qquad \rho_\pm = \sqrt{[y^2+(w\pm r_0)^2]}$  
\item  The $D_5$ solution has
$$ Z_x=Z_y=Z_0\qquad Z_\Omega = Z_0[{\rho_-^2\over \rho_-^2 + \rho_c^2}]^2$$

where $ \rho_c= {2gr_0\alpha'\over R^2}$ and $r_0=\pi\alpha' mN$

\item  The $NS5$ solution has
$$ Z_x=Z_\Omega=Z_0{\rho_-^2\over \rho_-^2 + \rho_c^2}
\qquad Z_y =
Z_0{\rho_-^2 + \rho_c^2\over \rho_-^2 }$$

where $ \rho_c= {2r_0\alpha'\over R^2}$ and $r_0=g\pi\alpha' mN$

{\bf Wilson loops }

\item To check whether the Wilson loops are of area law behavior 
we return to our  criterion stated in terms of
$$ f^2 = Z_x^{-1}\ \ \ \  \qquad \ \ \ \ g^2= {Z_y^{1/2} \over Z_x^{1/2}}. $$
so that $f^2$ and $g^2$ take   for the 
  $D_5$  and the $NS5$ cases the following values respectively, 
$ {\rho_+^2\rho_-^2\over R^2},  {\rho_-^2 + \rho_c^2\over
\rho_-^2},
{\rho_+^2+\rho_-^2\over \rho_-^2} ,  1 $

\item  {\bf Wilson loop of the $D_5$ case}

$f^2$ has a minimum with $f_{min}=0$, and $g^2=1$ does not diverge thus
{\it there is no confinement}.
In fact an explicit calculation shows that there is {\bf Screening} behavior.
\item  {\bf Wilson loop of the $NS5$ case}

$g^2$  diverges at $y=0, w=r_0$ where  $f(y=0,w=r_0)>0$
{\it there is  confinement}

\item 
{\bf Can we get in a similar manner the Wilson loop associated with the rest of
the possible vacua?}

\item Consider for example the case of $p$ $D_5$ branes that corresponds to
an $SU(p) \in SU(N)$ gauge theory\cite{KLSSY}.

\item Naively we expect the strings $(F1)$ to end on the $D_5$ branes
and hence to have screening. 

\item This is also the outcome of the use of the general theorem
  applies to the metric of the p $D_5$ branes ( $f(u_{min})=0$

\item However, from the field theory we know that quarks 
of the $SU(p)$ must confine. 

\item How do we resolve this contradiction?

\item  Recall that the world volume theory of the $p$ $D_5$ branes is
a $ SU(p)$ gauge theory.

\item A fundamental string ending on the $D_5$ branes  is a ``quark''
  of the  wv $SU(p)$ theory and thus can ``end''
 only provided

(i) if it is connected to an anti- quark string

(ii) if $p$ quarks combine to form a Baryon 


\item In this way of incorporating the wv theory we get that indeed
$(1,0)$ quarks confine and $(p,0)$ are screened.

\item 
For the case of unbroken $SU(p)$ 
 field theory also tells us that a magnetic monopoles  with
  charge $q={N\over p}$ has to be  screened. 
The naive use of the theorem tells us that any $D1$ string confines.

\item Again we have to use the full SUGRA  background.
Indeed  in the SUGRA picture there are
 D3 branes filling the $S^2$ sphere on
which the 5--branes are wrapped
 which behave as {\it baryon vertices.}
\item 
Those baryon vertices arise through the Hanany--Witten effect,
 when baryon vertex of the unperturbed ${\cal N} = 4$
theory, which is a 5--brane wrapping an $S^5$, contracts and moves
through the sphere.


\item  Now each D5
brane has a dissolved D3 charge of $q$
 the junction of a D3 ball with a wrapped D5
must support strings with total $D1$ charge of $q$ but $D1$s with different
charge cannot end and thus are confined.

\item  One can account for the various loops  associated with the other vacua.
\end{itemize}

\begin{center}
\fbox{
\bf Summary and open questions
}
\end{center}
\begin{itemize}
\item 
Indeed in all the stringy setups that supposed to be associated with 
confining dynamics we detect an area law Wilson loop. 
\item
Each  of the models suffers from certain problems and 
it seems that we have not found yet the optimal 
stringy laboratory to examine confinement. 
\item
There are indications that there is an attractive Luscher term.
To be contrasted with lattice simulations and phenomenology. 
\item 
The $N=1*$ case emphasized the fact that ( not only the metric)
but the full background affects the stringy loops. 

\end{itemize}

{\bf Acknowledgement}

Work supported in part by TMR program FMRX-CT96-0012.


\begin{thebibliography}{99}



\bibitem{Mal2} J. Maldacena, "Wilson loops in large $N$ field theories", Phys. Rev. Lett. 80 (1998) 4859-4862, hep-th/9803002.

\bibitem{Rey} S.-J. Rey and J. Yee,  ``Macroscopic strings as
    heavy quarks in the large N gauge theory and anti-de Sitter
    supergravity'', hep-th/9803001.




\bibitem{DGO} N. Drukker, D. Gross and H. Ooguri
 "Wilson Loops and Minimal Surfaces", hep-th/9904191 

\bibitem{KSS2}  Y. Kinar, E. Schreiber and J. Sonnenschein,
{\em ``$Q \bar{Q}$ Potential from Strings in Curved Spacetime -
  Classical Results''}, hep-th/9811192.
\bibitem{KSSW} Y. Kinar, E. Schreiber, J. Sonnenschein, N.Weiss
 ``Quantum fluctuations of Wilson loops from string models''
Nucl.Phys. B583 (2000) 76-104,
hep-th/9911123
\bibitem{Luscher}  M. Luscher, K. Symanzik, P. Weisz, 
"Anomalies of the free loop wave equation in the WKB approximation"
Nucl. Phys. B173 (1980) 365.

\bibitem{Arvis} J. F. Arvis Phys. Lett. 127B (1983) 106.

\bibitem{Alvarez}O. Alvarez, Phys. Rev. D24 (1981) 440. 


\bibitem{Witads2} E. Witten, "Anti-de Sitter Space, Thermal Phase
    Transition, and Confinement in Gauge Theories", hep-th/9803131.
\bibitem{BISY2} A. Brandhuber, N. Itzhaki, J. Sonnenschein, 
S. Yankielowicz, "Wilson Loops, Confinement,
 and Phase Transitions in Large N Gauge Theories from Supergravity", 
JHEP 9806 (1998) 001,hep-th/9803263
\bibitem{GriOle}  J. Greensite, P. Olesen,
 "Remarks on the Heavy Quark Potential in the Supergravity Approach", 
hep-th/9806235;
"Worldsheet Fluctuations and the Heavy Quark Potential 
in the AdS/CFT Approach", hep-th/9901057.

\bibitem{DorPer}
H. Dorn, V. D. Pershin `` Concavity of the $Q\bar Q$ potential in ${\cal N}=4$ super Yang-Mills gauge theory and
              AdS/CFT duality" hep-th/9906073

\bibitem{Bachas}C. Bachas,   Phys. Rv. D33 (1986) 2723.

 
\bibitem{MetTse} R. R. Metsaev, A. A. Tseytlin, "Type IIB superstring
  action in \adss background", hep-th/9805028.
 
\bibitem{Pesando} I. Pesando, "A $\kappa$ Gauge Fixed Type IIB
  Superstring Action on \adss", hep-th/9808020.

\bibitem{KalTse} R. Kallosh, A. A. Tseytlin, "Simplifying
  Superstring Action on \adss", hep-th/9808088.   











\bibitem{Polyakov} A.M. Polyakov,
{\em "The Wall of the Cave"}, hep-th/9809057.
\bibitem{KleTse1} I.R. Klebanov and A.A. Tseytlin,
{\em "D-Branes and Dual Gauge Theories in Type 0 Strings"}, hep-th/9811035.
 
\bibitem{Minahan} J.A. Minahan,
{\em ``Glueball Mass Spectra and Other Issues for Supergravity Duals
  of QCD Models''}, hep-th/9811156.
 
\bibitem{KleTse2} I.R. Klebanov and A.A. Tseytlin,
{\em "Asymptotic Freedom and Infrared Behavior in the Type 0 String
Approach to Gauge Theory"}, hep-th/9812089.

\bibitem{AFS}
 Adi Armoni, E.  Fuchs, J.  Sonnenschein
 ``Confinement in 4D Yang-Mills Theories from 
Non-Critical Type 0 String Theory''
 JHEP 9906 (1999) 027 hep-th/9903090
 
\bibitem{gppz}  L. Girardello, M. Petrini, M. Porrati and A. Zaffaroni,
   ``Confinement and Condensates Without Fine Tuning in
    Supergravity Duals of Gauge Theories'', hep-th/9903026.





\bibitem{PolStr}
J. Polchinski and M. J. strassler, ``The String Dual of a Confining Four 
Dimensional Gauge Theory'' hep-th/0003136

\bibitem{Myers}
R. C.  Myers, `` Dielectric-Branes'', JHEP 9912 (1999) 022, hep-th/9910053


\bibitem{KLSSY}
 Y. Kinar, A. Loewy, E. Schreiber, J. Sonnenschein, S. Yankielowicz
``Supergravity and Worldvolume Physics in the Dual Description 
of $N=1^*$ Theory'' hep-th/0008141




\end{thebibliography}
\end{document}